\documentclass[pdflatex,sn-basic,Numbered,referee]{sn-jnl}

\usepackage{graphicx}%
\usepackage{multirow}%
\usepackage{amsmath,amssymb,amsfonts}%
\usepackage{amsthm}%
\usepackage{mathrsfs}%
\usepackage[title]{appendix}%
\usepackage{xcolor}%
\usepackage{textcomp}%
\usepackage{manyfoot}%
\usepackage{booktabs}%
\usepackage{array}%
\usepackage[T1]{fontenc}%
\usepackage{lmodern}%

\begin{document}

\title[FRAME-C: knowledge-augmented deep learning for MEA classification]{FRAME-C: A knowledge-augmented deep learning pipeline for classifying multi-electrode array electrophysiological signals}

\author[1]{\fnm{Nisal} \sur{Ranasinghe}}
\author[2]{\fnm{Dzung} \sur{Do-Ha}}
\author[1]{\fnm{Chathura J.} \sur{Gamage}}
\author[2]{\fnm{Simon} \sur{Maksour}}
\author[1]{\fnm{Tamasha} \sur{Malepathirana}}
\author[1]{\fnm{Sachith} \sur{Seneviratne}}
\author*[2]{\fnm{Lezanne} \sur{Ooi}}\email{lezanne@uow.edu.au}
\author*[1]{\fnm{Saman} \sur{Halgamuge}}\email{saman@unimelb.edu.au}

\affil[1]{\orgdiv{AI, Optimization and Pattern Recognition Lab, Robotics Research Group, Department of Mechanical Engineering}, \orgname{The University of Melbourne}, \orgaddress{\city{Parkville}, \postcode{3010}, \state{VIC}, \country{Australia}}}

\affil[2]{\orgdiv{School of Science and Molecular Horizons}, \orgname{University of Wollongong}, \orgaddress{\city{Wollongong}, \postcode{2522}, \state{NSW}, \country{Australia}}}

\abstract{Amyotrophic lateral sclerosis (ALS) is a fatal neurodegenerative disorder characterized by motor neuron degeneration, with changes to neuronal excitability being an early phenotype for ALS identified in patients and animal models. Recent advancements in induced pluripotent stem cell (iPSC) technology have enabled the generation of human iPSC-derived neuronal cultures, which, when combined with multi-electrode array (MEA) electrophysiology, provide rich spatial and temporal electrophysiological data. Traditionally, MEA data is analyzed using handcrafted features based on potentially imperfect domain knowledge, which while useful may not fully capture all the relevant characteristics inherent in the MEA data. Machine learning, in particular deep learning has the potential to automatically learn relevant characteristics (features) from raw data, without solely relying on handcrafted feature extraction. However, handcrafted features remain critical for encoding domain knowledge and improving model interpretability, especially in scenarios with limited or noisy data, as is often the case in most experimental studies. This study introduces FRAME-C, a knowledge-augmented machine learning pipeline that combines domain knowledge, raw spike waveform data, and deep learning techniques to classify MEA signals and identify ALS-specific phenotypes. FRAME-C leverages deep learning to extract important features from spike waveforms, while also incorporating handcrafted features such as spike amplitude, inter-spike interval, and spike duration, thus preserving key spatial and temporal information. We validated FRAME-C on both simulated and real-world MEA data from human iPSC-derived neuronal cultures, demonstrating its superior performance compared to existing methods for MEA classification. FRAME-C performs better, showing more than a 36\% relative improvement on real-world data and up to 46\% relative improvement on simulated data in terms of the test accuracy. Moreover, we show that FRAME-C can be used to evaluate the importance of each of the handcrafted features, thereby contributing to the interpretation of the classification results. Our analysis using permutation feature importances revealed that inter-spike interval and spike amplitude are the most important features for classification, suggesting that these features are key phenotypes of ALS iPSC-derived motor neurons, compared to controls.}


\maketitle

\section{Introduction}
\label{sec:introduction}
Amyotrophic lateral sclerosis (ALS) is a fatal neurodegenerative disease that leads to a progressive loss of motor neurons. At the onset of ALS, symptoms may include limb weakness and difficulty in swallowing. However, the disease commonly progresses towards paralysis and respiratory failure within three to five years \cite{brown_amyotrophic_2017}.

A small portion of ALS patients (5 - 10\%) have a familial pattern of inheritance (fALS)  and can be linked to a family history of ALS. However, the majority (90 - 95\%) are sporadic (sALS) and do not have any known family history. Many studies have been conducted on fALS to identify genes that contribute to the disease \cite{rosen_mutations_1993, renton_hexanucleotide_2011, dejesus-hernandez_expanded_2011, sreedharan_tdp-43_2008}, and more than 40 genes have been linked to fALS \cite{chen_genetics_2013}, with many more believed to be unknown. While fALS has been well studied using mouse models, this is not possible for sALS since the underlying cause(s) are more complex.

Human induced pluripotent stem cell technology (iPSC) provides a platform to study motor neuron diseases \cite{takahashi_induction_2006, takahashi_decade_2016}. They allow the development of neuronal cultures that mimic the morphology, gene expression and function of neurons. Since the neuron cultures are derived from human donor stem cells, they are a powerful model for studying both fALS and sALS.

Changes in excitability throughout disease progression are well-documented in ALS \cite{do-ha_impairments_2018, wainger_intrinsic_2014, buskila_dynamic_2019}, and patient iPSC-derived cultures provide an opportunity to better study the excitability of ALS donor-derived neurons.

Multi-electrode arrays (MEA) allow the recording of electrophysiological signals within \textit{in vitro} neuronal cultures. They consist of multiple electrodes that can simultaneously detect changes in extracellular voltage when neurons fire action potentials. Since each electrode records the activity from different regions of the neuronal culture with high resolution, MEAs provide rich information about both the timing (temporal resolution) and location (spatial resolution) of neural activity. When these MEAs are used with iPSC-derived neuronal cultures, they provide a powerful platform to study disease specific phenotypes in neuronal activity \cite{mossink_human_2021}. 

However, the richness of information in MEA data introduces many challenges in understanding and analyzing the data. The high dimensionality of MEA recordings makes it challenging to analyze using manual or semi-automated processes. Traditionally, MEA signals are analyzed by extracting predefined features from the recording, and performing statistical analysis on these features \cite{bryson_classification_2022}. ``Spikes''--rapid changes in electrical activity that indicate neuron firing—are considered the most informative aspect of MEA data. Traditional methods focus on spike timing and firing rates, which are believed to capture the majority of the useful phenotypes that describe cell behaviour and network dynamics. However, relying on predefined features limits the ability to fully leverage the detailed spatial and temporal information present in MEA recordings. The advancements of machine learning, in particular deep learning provide a powerful alternative to learn the complex relationships within these high dimensional MEA datasets \cite{zhao_deep_2019, buccino_combining_2018, malepathirana_visualization_2024}. Deep learning methods generally do not rely on handcrafted features, since they can automatically extract latent features from unstructured raw data such as images, time series and text \cite{dong_survey_2021}. These models have achieved state-of-the-art results in various domains, demonstrating their potential for analyzing complex MEA signals.

In this work, we introduce Feature-based Raw data Augmentation for Multi-electrode array recorded Electrophysiological signal Classification (FRAME-C), a machine learning pipeline for classifying MEA signals based on electrophysiological phenotypes. FRAME-C combines domain knowledge with deep learning by integrating handcrafted features into the deep learning pipeline. We validate this pipeline on both simulated MEA data and \textit{in vitro} MEA data obtained from human iPSC-derived neuronal cultures. To the best of our knowledge, this study represents the first application of deep learning methods to classify MEA recordings in the context of ALS research.

The main contributions of this paper are:
\begin{enumerate}
    \item A knowledge augmentation method to integrate features derived from domain knowledge into deep learning methods for more accurate classification.
    \item FRAME-C: a machine learning pipeline that uses knowledge augmentation, data augmentation and deep learning to classify MEA signals.
    \item A detailed analysis of how each component of the pipeline affects the classification performance of FRAME-C.
    \item Feature importance analysis of the trained model, uncovering electrophysiological markers of ALS.
\end{enumerate}

In Section \ref{sec:related_work}, we review the related literature on electrophysiological analysis, with a particular focus on MEA recordings and machine learning. Then, in Section \ref{sec:dataset} we introduce the datasets that we use for evaluating FRAME-C, including a simulated dataset and real-world MEA data recorded from human iPSC-derived neuronal cultures. In Section \ref{sec:methods} we introduce FRAME-C and discuss the components of the ML pipeline. The results of the conducted experiments are then presented in Section \ref{sec:results} followed by a summary of the work, limitations and future research directions in Section \ref{sec:conclusion}. 

\section{Related work}
\label{sec:related_work}

Machine learning (ML) methods have been used to perform predictive analytics using electrophysiological signals such as Electroencephalography (EEG) \cite{hosseini_review_2021, li_temporal-spectral_2024}, Electrocardiography (ECG) \cite{al-zaiti_machine_2023} and Electromyography (EMG) \cite{yousif_assessment_2019}. However, there are only a few studies which explore the use of ML techniques in the study of MEA electrophysiology. 

Traditionally, MEA data is analyzed by extracting features from action potentials and performing statistical analysis on them. Some commonly used features include mean firing rate, mean bursting rate, network bursting rate and burst duration. Such analysis of MEA data has been shown to be a powerful tool in studying disease specific phenotypes \cite{mossink_human_2021}. However, these methods do not make full use of the rich spatial and temporal data in MEA recordings, since they can only study the pre-defined features that were extracted from the signals.

Buccino et al. \cite{buccino_combining_2018} proposed a deep learning pipeline for classification and localization of neurons using extracellular simulations. In this work, simulated waveforms are used to train a convolutional neural network (CNN) to predict the type and location of a neuron based on their extracellular action potentials (EAP). Features extracted from the EAPs are used to create 2D feature maps, with each point in the feature map corresponding to the feature value recorded on a given electrode. Multiple feature maps are stacked to create a 3D feature map which is then fed into the CNN for training. However, since each training sample corresponds to a single action potential, the network cannot learn from the rich temporal information in the data. Moreover, the method is only applied to simulated EAPs, and has not been shown to be effective on real MEA data. 

Zhao et al. proposed a convolutional neural network (CNN) to classify \textit{in vitro} MEA recordings \cite{zhao_deep_2019}. The authors apply this CNN to classify different genotypes using mouse iPSC-derived neurons and human iPSC-derived neurons with Williams syndrome. Before passing into the CNN, spike detection is performed on each MEA recording to transform the signal into a binary signal where 1 represents a spike and 0 represents a lack of a spike. Though this network is shown to perform better than simple machine learning methods such as logistic regression, it can be argued that the spike detection step removes a lot of useful information from the signal, since the model only sees whether the neuron fired or not, and does not see any information related to the shapes of the spikes. 

Deep learning has also been applied directly to extracellular spike recordings to improve signal quality. Xu et al. proposed a spike-denoising model combining a convolutional autoencoder with bidirectional LSTM and attention layers, which outperformed traditional denoising methods on both synthetic spikes and recordings from cultured cortical neurons \cite{xu_enhancing_2025}. However, such approaches target signal pre-processing rather than the classification of disease-related phenotypes from MEA recordings.

Though most of the current work on MEA classification relies on handcrafted features based on domain knowledge, deep learning methods in general do not require handcrafted features. The hallmark of deep learning is its ability to automatically learn features from data. By leveraging the spatial and temporal richness of MEA recordings, deep learning approaches have the potential to uncover patterns and interactions that may not be apparent through traditional feature extraction methods. However, handcrafted features still play a crucial role, especially in scenarios where data is limited or the interpretability of the model is important. These features, derived from domain expertise, can provide insights into specific characteristics of the data that may guide the model's learning process. Additionally, combining handcrafted features with deep learning models can often enhance performance by incorporating prior knowledge, ensuring that the model captures both known and latent patterns in the data \cite{cui_knowledge-augmented_2023, gamage_knowledge_2026}. In neurodegenerative disease detection more broadly, fusing complementary feature sources is a common strategy for improving performance, whereas the explainability of such fused models remains comparatively under-explored \cite{viswan_multimodal_2026}. The integration of domain knowledge with deep learning is known as knowledge augmentation \cite{cui_knowledge-augmented_2023, bogahawatte2026model}.

\section{Dataset} 
\label{sec:dataset}

We evaluate our ML pipeline using two datasets.
\begin{itemize}
    \item MEA data recorded from human iPSC derived neuronal cultures 
    \item MEA data simulated using layer 5 cortical neuronal models
\end{itemize}

\subsection{Human iPSC-derived neuronal culture data}

\subsubsection{Motor neuron differentiation}

The study was conducted according to the National Statement on Ethical Conduct in Human Research and approved by the University of Wollongong Human Research Ethics Committee (approval numbers 13/272 and 19/380). 


Four sporadic ALS and four age- and sex-matched donor iPSC lines were utilised in this study (Table \ref{tab:cell_lines}). sALS 2 and Healthy Control 2 are identical twins. iPSCs were maintained as previously described in Maksour et al., 2024 \cite{maksour_alzheimers_2024}. Motor neuron progenitors (MNPs) were differentiated as previously described in Robinson et al., 2026 \cite{robinson_als-ftd-linked_2026}, and cryopreserved prior to terminal motor neuron differentiation. 48-well CytoView MEA plates were coated with 100 $\mu$g/mL poly-D-lysine in PBS for 1 h at room temperature, followed by three washes with phosphate-buffered saline (PBS). Plates were then coated with Matrigel overnight at 4\,$^{\circ}$C. MNPs were seeded as single cells onto the coated MEA plates at a density of 40,000 cells/well in motor neuron media (BrainPhys, 1x N2 supplement, 1 x B27 supplement, 1x NEAA) supplemented with 10 ng/mL BDNF, GDNF, IGF, 500 ng/mL DB-cAMP, 200 ng/mL L-AA, 0.5 $\mu$M retinoic acid, 0.1 $\mu$M purmorphamine and 10 $\mu$M DAPT, with half media changes every other day for the duration of MEA recording. Following 10 days of maturation, retinoic acid, purmorphamine and DAPT were removed from motor neuron media. Every second media change was supplemented with 2 $\mu$g/mL laminin to promote cell adhesion.

\begin{table}[htbp]
\caption{Donor demographics of the iPSC lines used in this study}\label{tab:cell_lines}
\begin{tabular}{@{}lll@{}}
\toprule
Cell line & Sex & Age (years) \\
\midrule
sALS 1 & M & 51 \\
Healthy Control 1 & M & 50 \\
sALS 2 & M & 46 \\
Healthy Control 2 & M & 46 \\
sALS 3 & M & 50 \\
Healthy Control 3 & M & 52 \\
sALS 4 & F & 82 \\
Healthy Control 4 & F & 75 \\
\botrule
\end{tabular}
\footnotetext{sALS: sporadic amyotrophic lateral sclerosis; M: male; F: female. sALS 2 and Healthy Control 2 are identical twins.}
\end{table}  

\subsubsection{Microelectrode array measurements}

Motor neuron cultures were maintained for 1 week prior to recording spontaneous neuronal network activity using the Axion Maestro system. Recordings were performed at 37\,$^{\circ}$C and 5\% CO$_2$. Spontaneous neuronal network activity was recorded 3-4 times a week prior to media changes for 9 weeks. Electrophysiological data was acquired through the AxIS software (version 2.5) at a sampling rate of 12.5 kHz and passed through a high and low pass Butterworth filter (200 Hz -- 3 kHz). For each cell line, three independent differentiation replicates were analysed, each with two technical replicates (wells per plate).

\subsubsection{Dataset exploration}

\begin{figure}[htbp]
  \centering
  \includegraphics[width=\linewidth]{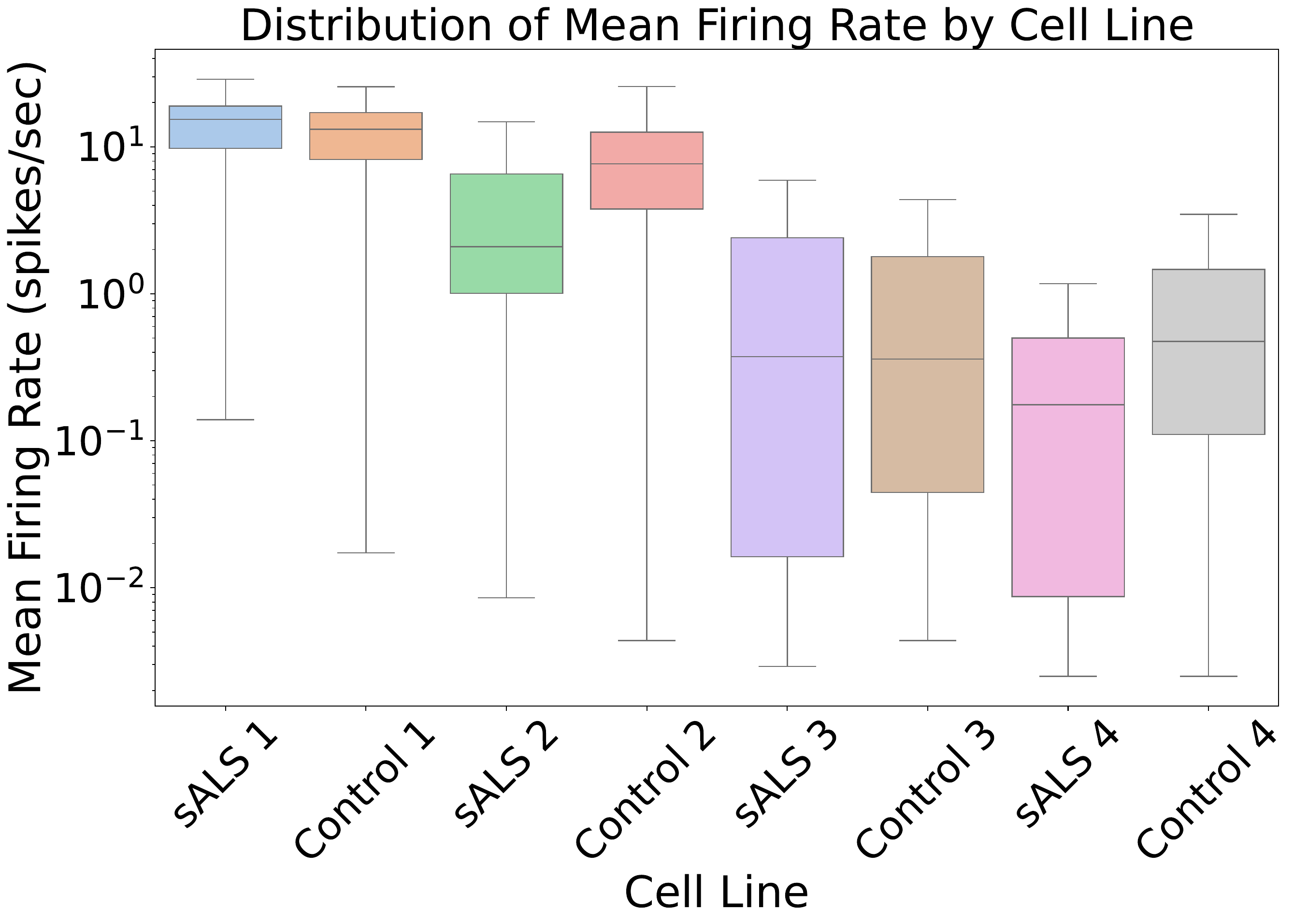}
  \caption{\textbf{Distribution of mean firing rates (MFRs) of the MEA recordings for each cell line.} MFRs of all the MEA signals recorded from human iPSC-derived neuronal cultures, shown per cell line on a log scale. Boxes indicate the interquartile range (25th--75th percentiles), the centre line indicates the median, and whiskers extend to the most extreme values within 1.5 times the interquartile range. Data were obtained from three differentiation replicates per cell line, each with two technical replicates (wells).}
  \label{fig:mfr_by_cell_line}
\end{figure}

We performed exploratory analysis to further understand the dataset. The threshold crossing algorithm is used to detect action potentials in the recordings, which are then used to calculate the mean firing rates of the recordings. The mean firing rate $MFR$ is defined as follows.

\begin{equation}
    MFR = \frac{\sum_{i=1}^{N} spikes_i}{N \cdot T}
\end{equation}

where $N$ is the number of electrodes, $spikes_i$ is the number of spikes detected in the $i$th electrode, and $T$ is the duration of the recording. The MFRs of all the MEA signals are illustrated using a box plot in Figure \ref{fig:mfr_by_cell_line}. Despite the analysis of three differentiation replicates and two technical replicates per cell line, the MFRs of the last four cell lines were markedly lower than those of the first four cell lines.

Despite these observable differences, we incorporated all eight cell lines into our primary analysis to enable broader, disease-specific statements that encompass a wide range of biological variability. However, to rigorously assess the proposed method's performance on high-quality data, we also conducted a separate evaluation using only the first four cell lines, excluding the lower-activity cultures. 

Furthermore, we evaluated our pipeline exclusively on the sALS 2 and Control 2 cell lines, which were derived from identical twins. In general, training a deep-learning model on data from only two subjects is not appropriate for binary classification because the model may learn subject-specific characteristics unrelated to ALS phenotypes. However, in the case of identical twins, genome-related differences are more limited than for unrelated healthy controls, reducing the risk that the model learns subject-specific traits unrelated to the sALS phenotype.

\subsection{Simulated data}

To further evaluate the proposed pipeline, we simulated MEA recordings using the MEArec simulator \cite{buccino_mearec_2021}. MEArec is a fast, open-source Python based simulator to simulate MEA recordings. It uses the NEURON simulation environment \cite{carnevale_neuron_2006} for intracellular simulation which is then used for an extracellular simulation using the LFPy package \cite{linden_lfpy_2014}. This produces the extracellular action potentials generated at each electrode location.

For simulating the MEA recordings, we used layer 5 (L5) cortical neuronal cell models from the Neocortical Microcircuit Portal \cite{ramaswamy_neocortical_2015}. This consists of 13 different cell models, some of which are inhibitory cells and the rest of which are excitatory cells. Each recording is simulated for 300 s with a sampling rate of 12.5 kHz. To better represent noisy MEA data, we added Gaussian noise to each recording of the simulated dataset.

Since we are interested in evaluating the performance of the proposed ML pipeline in its ability to distinguish between cell characteristics, two MEA recording datasets were simulated for each parameter configuration. 

\begin{itemize}
    \item Dataset 1 (Label 0): simulated with 10 excitatory cells
    \item Dataset 2 (Label 1): simulated with 10 inhibitory cells
\end{itemize}

\section{Methods}
\label{sec:methods}

\begin{figure}[htbp]
  \centering
  \includegraphics[width=\linewidth]{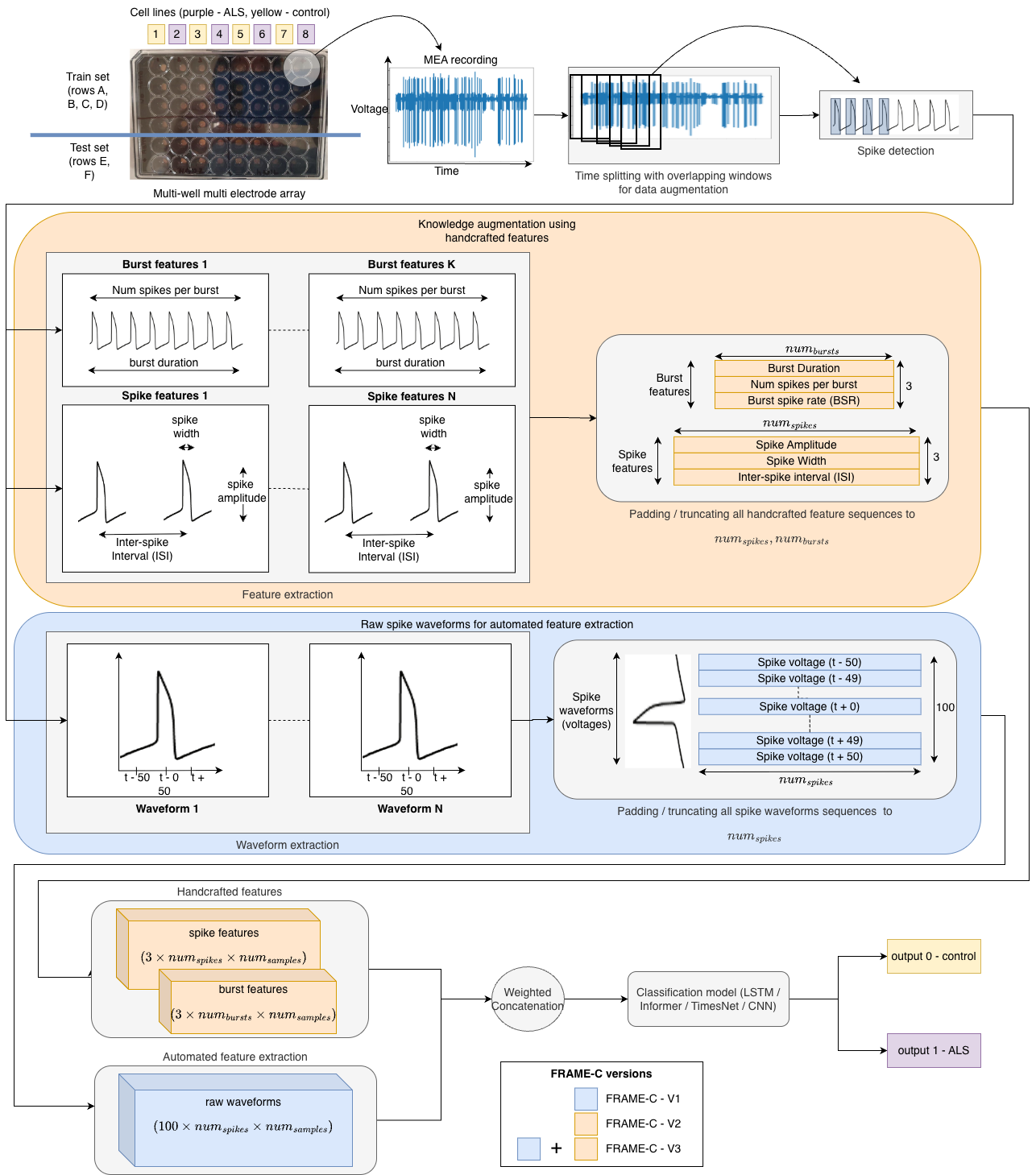}
  \caption{\textbf{Architecture of the proposed FRAME-C pipeline.} The raw spike waveforms are augmented using handcrafted features based on domain knowledge. The sequences of spikes are truncated/padded to a fixed length of $len_{spikes}$, while the sequences of bursts are truncated/padded to a fixed length of $len_{bursts}$.}
  \label{fig:ml_pipeline}
\end{figure}

Within this section, we introduce the ML pipeline and explain each component in detail. We first explain the pre-processing pipeline which extracts features and creates the dataset, and then the deep learning architecture that we use for classification. A detailed figure of the architecture of the ML pipeline is given in Figure \ref{fig:ml_pipeline}.

\subsection{Noise filtering and time split}

We first apply a 4\textsuperscript{th} order Butterworth filter with cut-off frequencies of 0.3 and 2 kHz. Each signal is 300 s in length, and is sampled at 12.5 kHz. To increase the number of data samples, and to decrease the dimensionality of each sample, we split each signal into multiple smaller signals. Each time split signal is then considered as a single data sample. 

\subsection{Spike waveform extraction}

The next step in the pipeline involves the detection of spikes and extracting the corresponding spike waveforms. Action potential (AP) detection, or spike detection, was performed using the threshold crossing algorithm, with a threshold of 5 times the standard deviation of the noise level. The spike waveforms are captured by extracting 50 samples before and after the detected peak of the spike, yielding waveforms of 100 samples per spike. Each waveform is represented by a 100-dimensional feature vector, corresponding to the voltage values of the spike over these 100 sample points. The full spike waveform captures the finer temporal details of the spike, which may not be captured by extracting handcrafted features which is often conducted in traditional analysis. We expect the deep learning model to capture useful phenotypes of the spike using these raw waveforms.

\subsection{Knowledge augmentation using spike features}

Although deep learning models have been shown to be very successful in learning latent features from raw data, incorporating domain-specific handcrafted features can further enhance the model's performance \cite{cui_knowledge-augmented_2023}. We further augment the dataset using domain knowledge by extracting features from the detected spikes and adding them to the spike waveform data. These features encode expert knowledge that may be challenging for the model to learn autonomously, especially from limited and noisy data.
By adding these features, we also aim to improve the model's interpretability. Having interpretable features allows us to later perform feature importance experiments to further understand the phenotypes that were useful for the model in classifying the MEA recordings. Moreover, handcrafted features can capture biologically relevant patterns that are not immediately obvious in raw data, that can lead to better generalization.

From each spike, the following features are extracted:

\subsubsection{Spike amplitude}
This is the amplitude of a given AP. This is measured as the difference between the baseline noise level and the peak voltage value of the detected AP. Although the AP amplitude is not a commonly used feature in MEA analysis in ALS studies, it has been identified as a key feature in studies of other neurodegenerative diseases such as Alzheimer's disease \cite{wang_opposite_2009}.

\subsubsection{Inter-spike interval (ISI)} This is the time interval between the AP and the next AP in the MEA signal. The ISI is a common phenotype used in MEA studies \cite{mossink_human_2021}.

\subsubsection{Spike duration / spike width} This is the width of the spike at 50\% of the spike amplitude. We first identify the two points at which the MEA voltage is 50\% of the maximum absolute voltage of the AP. The time interval between these two points is the spike duration, or the spike width \cite{weir_comparison_2015}.

Once the APs are detected, we further augment the dataset by detecting bursts, and corresponding burst related features. Bursts are regions in the MEA recording where a number of APs are fired within short intervals of each other, indicating regions of heightened excitability. Bursts that occur within a single channel of an MEA recording are known as single channel bursts. We detect single channel bursts by identifying consecutive APs with ISI values below a given $min_{ISI}$ threshold. Moreover, we recognize such APs as bursts, only if they have at least $min_{spikes}$ number of APs. We set $min_{ISI}$ to 8 ms and $min_{spikes}$ to 4. Once the bursts are detected, we extract the following features:

\subsubsection{Burst duration} This is the time between the first and last AP of a given burst.

\subsubsection{Number of spikes} This is the total number of spikes in the burst.

\subsubsection{Burst spike rate (BSR)} This is the rate of firing of spikes within each burst. The BSR is calculated as $Burst\ duration / Number\ of\ spikes$.

Apart from spikes and single channel bursts, another commonly studied phenomenon in MEA recordings is network bursts \cite{mendis_use_2016}. A network burst occurs when multiple single channel bursts are recorded simultaneously across multiple electrodes. This is an indicator of network activity in the neuronal culture, where multiple groups of neurons fire at once. In this study, we do not explicitly detect network bursts and extract network burst related features. This is because we do not remove the temporal information of the data when feeding it into the deep learning model. Therefore, we expect the deep learning model to capture any relevant network behaviour when training. 

\subsection{Creating feature sequences}

Once spikes are detected, and the spike waveforms and features are extracted, feature sequences are created for each recording split. Since the spike waveform consists of 100 data points, and there are three extracted spike features, this results in a 103-dimensional feature vector for each spike. This results in a feature sequence created for each recording split. When burst features are also included, we create a similar 3-dimensional feature sequence using just the extracted burst features. For example, if a recording split contains 2000 spikes, the spike feature sequence would be of shape $103 \times 2000$. Similarly, for 200 bursts, the burst feature sequence would be of shape $3 \times 200$.

To ensure uniformity across recordings with varying numbers of spikes and bursts, padding or truncation is applied, producing feature sequences of dimensions $103 \times n$ and $3 \times m$ where $n$ is the fixed length of the spike sequences and $m$ is the fixed length of the burst sequences. This approach preserves both temporal and spatial information while substantially reducing the dimensionality of the dataset, making it more efficient for subsequent analyses. By integrating both spike features and the waveform, the method achieves a balance between computational efficiency and the retention of biologically relevant signal characteristics, enhancing the overall quality of data representation for downstream tasks.

\subsection{Deep learning model}

Since the preprocessed dataset comprises a sequence of features, we first apply projected concatenation to fuse the raw waveforms and handcrafted features into a unified representation. This fusion mechanism uses learnable weights that enable the model to dynamically adjust the importance of each data modality, optimizing the balance between raw signal information and domain expert knowledge. We then use deep learning architectures that are successful in learning relationships within sequences of data. We perform experiments with several deep learning architectures and compare their performance in accurately predicting the correct cell type. 

\subsubsection{Long short term memory (LSTM) networks}
Recurrent neural networks (RNN) are a class of neural networks that are specifically designed to learn relationships in sequential data. RNNs have been applied in many fields such as natural language processing, time series analysis and even neuroscience \cite{barak_recurrent_2017}. The long short term memory (LSTM) network is a type of RNN that mitigates many of the problems of the vanilla RNN, allowing it to learn long term dependencies in long sequences of data. However, even LSTMs may fail in learning very long term dependencies in long sequences of data.

\subsubsection{Transformers and self-attention}

Transformers, a more recent and powerful architecture, have revolutionized sequence modeling by addressing some of the limitations of RNNs and LSTMs. Unlike recurrent models, transformers process entire sequences in parallel using "self-attention" mechanisms, enabling them to capture long-range dependencies  efficiently. This architecture has been highly successful in tasks like natural language processing, protein structure prediction, and image recognition \cite{vaswani_attention_2017, dosovitskiy_image_2020}. Its ability to model global context without relying on sequential computation makes it a strong candidate for tasks involving complex sequential data. Although the transformer architecture was originally proposed for natural language processing tasks, it has rapidly been adopted in many fields which require the processing of sequential data, including time series data analysis. 

Although transformers have mostly replaced LSTMs as the state-of-the-art architecture in natural language processing, and have been adopted in many other fields, there is still evidence that RNNs and LSTMs still outperform transformer based methods in certain applications \cite{buestan-andrade_comparison_2023}. Transformers are computationally expensive and have a large number of parameters, and are therefore prone to overfitting, especially when trained on small datasets.

The power of transformers comes at a high computational cost. The computational complexity of traditional self-attention scales quadratically with $n$, where $n$ is the length of the sequence \cite{keles_computational_2022}. This makes it computationally infeasible for early transformer based architectures to learn from long time series data. However, more recent works have proposed improved transformer architectures that have lower computational complexities, allowing them to be trained on very long time series data. We are particularly interested in this type of model since the MEA data can have long sequence lengths. The time series library \cite{wang_deep_2024}, a recent benchmark for deep learning models specialized in time series data compares many state of the art models on various tasks. In our experiments, we use two deep learning models that are shown to achieve state-of-the-art results in time series classification.

\textbf{Informer}. The Informer architecture is designed to overcome the limitations of traditional transformers in modeling long sequence data, such as time series \cite{zhou_informer_2021}. By introducing a probSparse self-attention mechanism, Informer reduces the computational complexity of the self-attention operation from $O(n^2)$ to $O(n \log n)$, where $n$ is the sequence length. This sparsity mechanism selects only the most informative queries for attention calculations, allowing Informer to process long sequences efficiently without sacrificing performance. Additionally, it incorporates a distillation module that hierarchically compresses the input sequence, retaining only essential features and reducing redundancy. These innovations make Informer a highly efficient and scalable architecture, well-suited for tasks such as time series forecasting and anomaly detection. Informer has demonstrated strong performance on various long sequence tasks, showcasing its utility in handling time series data with high dimensionality.

\textbf{TimesNet}. TimesNet is an architecture developed to address the challenges of learning from time series data with diverse temporal patterns \cite{wu_timesnet_2022}. Instead of relying on self-attention mechanisms, TimesNet employs a multi-scale convolutional framework to capture both localized and global temporal features. This approach allows it to efficiently model short-term fluctuations and long-term trends, which are critical for tasks like forecasting and classification. TimesNet also incorporates domain-specific priors, such as seasonal and periodic components, directly into its architecture, enhancing its ability to handle time series with strong temporal structures. Its modular design ensures scalability and interpretability, making it a practical choice for real-world applications. Benchmark evaluations have shown that TimesNet achieves state-of-the-art results on various time series tasks, offering a compelling alternative to transformer-based methods.

\subsection{Data Augmentation}

Deep learning models perform better when they are trained on large amounts of data. Although it is possible to generate as much data as needed when simulating recordings, collecting more real-world data in the form of MEA recordings is not always feasible. This is a common challenge with many deep learning models which rely on labelled datasets for training. Data augmentation is a technique used to artificially create more data from existing data, so that the ML model can be trained on a larger dataset.

As discussed earlier, we perform a time splitting of the MEA recording, since each recording is 300 seconds long. To further augment the size of the data, we perform a time split with a sliding window. The sliding window approach involves dividing the 300-second recordings into overlapping segments, where each window has a fixed duration (e.g., 10 seconds) and slides forward by a predetermined step size (e.g., 5 seconds). This method ensures that the entire recording is covered multiple times, creating overlapping segments that capture temporal continuity while enriching the dataset with more training samples. The overlap allows the model to learn subtle temporal patterns by exposing it to slightly shifted versions of the same data. This augmentation technique is particularly effective in retaining the contextual dynamics of the original recordings, thereby improving the robustness and generalization of the model trained on these enhanced datasets.

\subsection{Batch effects observed}

Batch effects refer to variations in a dataset caused by differences in technical or experimental conditions rather than biological differences. These variations, known as "non-biological variations," pose challenges for analyzing data with multiple batches. Factors that contribute to batch effects include differences in electrode sensitivity, sample preparation, and environmental conditions. Since ALS MEA data is generated using a multi-well MEA system, some degree of batch effects is expected in each well.

To investigate batch effects in the ALS dataset, we applied Uniform Manifold Approximation and Projection (UMAP) \cite{mcinnes_umap_2018} to reduce the dataset's dimensionality and visualize the results in two dimensions. Each feature sequence was first flattened into a $k \times 1$ array, where $k$ represents the total number of features. UMAP then reduced the data to two dimensions, which were visualized as a scatter plot. To assess the presence of batch effects, the points in the scatter plot were colored by the well number, allowing us to identify clusters based on the originating MEA wells. In the absence of batch effects, colors in the scatter plot would be randomly distributed, with no visible clusters.

\begin{figure}[htbp]
  \centering
  \includegraphics[width=\linewidth]{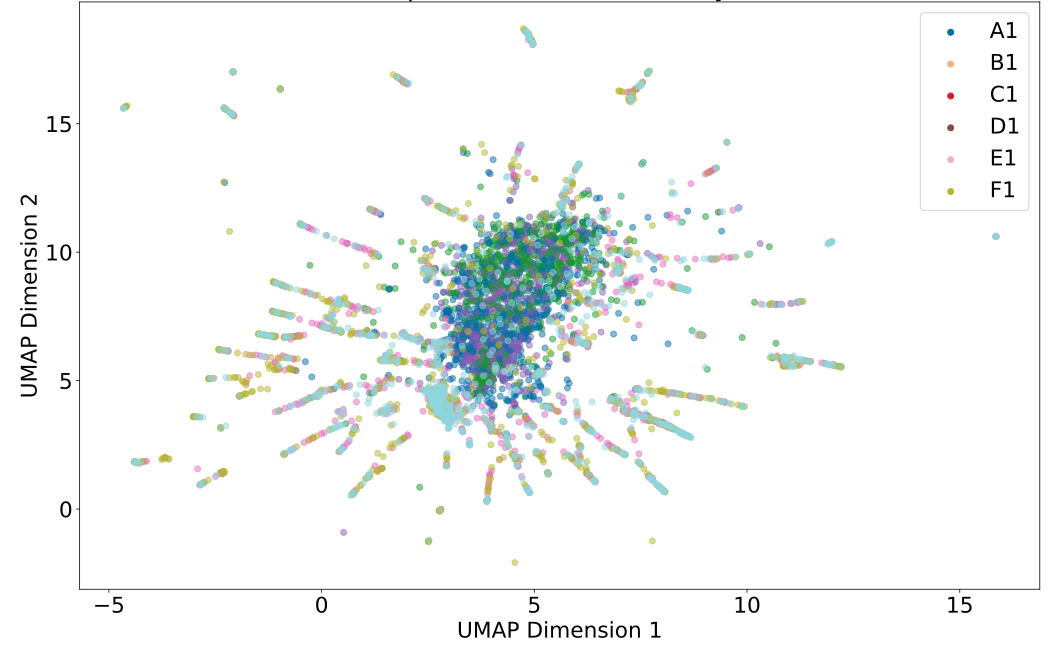}
  \caption{\textbf{UMAP visualization of spike features for sALS 1.} The projection is colored by the well number from which the MEA signal was recorded. The emergence of distinct clusters indicates the presence of batch effects within the dataset.}
  \label{fig:umap_batch_effects}
\end{figure}

The UMAP visualization of data recorded at 36 days of maturation is shown in Figure \ref{fig:umap_batch_effects}. To simplify the visualization, we present only the wells corresponding to a single cell line, as including all wells would result in an overly complex plot. The figure reveals clustering patterns associated with the originating wells, indicating the presence of batch effects in the dataset. Notably, similar clustering patterns were observed across datasets from all wells and at different maturation stages, further confirming the widespread presence of batch effects in the data.

\subsection{Well-wise train-test split}

When creating a train-test split for training and evaluating the model, we considered two strategies: (1) random split and (2) well-wise split. The random split involves randomly selecting a percentage of data samples for training and the rest for testing. However, this would mean that the model would see data from the same well during both training and testing. This gives rise to the possibility of the model relying on well-specific patterns (created due to batch effects) when performing the classification, leading to an overestimation of performance. To account for the batch effects observed in the ALS dataset, we instead chose a well-wise train-test split, meaning that data samples from the same well are kept entirely within either the training or testing set. This ensures that the model does not see data from the same well during both training and testing. By splitting the data well-wise, we ensure that the evaluation reflects the model’s ability to generalize to data from wells it has not seen before, providing a more accurate assessment of its performance.

In our experiments, we used well rows A, B, C and D for training and rows E and F for testing.

\subsection{Model Evaluation}

The real-world MEA dataset comprises an equal number of ALS and control cell lines, resulting in a balanced dataset. This balance justifies the use of accuracy as the evaluation metric, defined as the proportion of correctly classified samples across both classes. By ensuring equal representation of both classes, the dataset avoids class imbalance bias, providing a clear and reliable measure of the model's predictive performance. Accuracy is defined as follows.

\begin{equation}
   \mathrm{accuracy} = \frac{\text{number of correctly classified samples}}{\text{total number of samples}} 
\end{equation}

The CNN-based method proposed by Zhao et al. employs a different time-splitting strategy than the one used in our proposed ML pipeline. Consequently, directly comparing accuracy values does not provide a like-for-like comparison, as the methods are evaluated on recording splits of differing sizes. To address this, we use a voting method to generate predictions for full recordings (300 seconds each). Specifically, predictions are generated for each time split within a recording, and the classification most frequently occurring across the splits is assigned to the entire recording. For instance, if a recording contains 20 time splits, of which 15 are classified as ALS and 5 as healthy, the overall classification for that recording would be ALS.

\section{Results and Discussion}
\label{sec:results}

The human iPSC-derived neuronal culture dataset consists of recordings taken on 40 different days, spanning maturation periods from 9 to 78 days. For each dataset, independent classification models were trained and tested, with test accuracies recorded for comparison. We evaluate 4 variations of FRAME-C, where each variation uses a different deep learning architecture for classification. These are (1) Informer \cite{zhou_informer_2021}, (2) TimesNet \cite{wu_timesnet_2022}, (3) LSTM and (4) CNN. We compare these against the CNN-based MEA classification pipeline proposed by Zhao et al. \cite{zhao_deep_2019}. To the best of our knowledge, this pipeline is the only other deep learning based pipeline for MEA classification that has been evaluated on \textit{in vitro} MEA data. To obtain a fair comparison, we perform data augmentation for all methods by a factor of 10, including the pipeline proposed by Zhao et al. 

To comprehensively evaluate the effect of the FRAME-C pipeline and knowledge augmentation, we conduct multiple experiments for each classification model and present the results in Table \ref{tab:results_all_8}. We report the mean test accuracies achieved by each method, calculated by averaging the test accuracy values across all 40 datasets. Additionally, Table \ref{tab:results_filtered} demonstrates the improved accuracies observed when unreliable data samples are removed from the analysis. Unreliable samples were identified using the distribution of mean firing rates across wells as discussed in Figure \ref{fig:mfr_by_cell_line}. Finally, Table \ref{tab:results_twins} presents the classification performance when the models are applied to differentiate between two cell lines derived from identical twins, distinguishing between the ALS-affected subject and the healthy control.

\begin{table}[htbp]
\caption{Effect of handcrafted features on test accuracy using all eight cell lines}\label{tab:results_all_8}
\footnotesize\setlength{\tabcolsep}{3pt}
\begin{tabular*}{\textwidth}{@{\extracolsep\fill}>{\raggedright\arraybackslash}p{2.0cm}>{\raggedright\arraybackslash}p{1.9cm}>{\raggedright\arraybackslash}p{2.2cm}cccc@{}}
\toprule
Classification pipeline & Automated feature extraction & Handcrafted features (domain knowledge) & \multicolumn{4}{@{}c@{}}{Deep learning classifier} \\\cmidrule{4-7}
 & & & Informer & TimesNet & CNN & LSTM \\
\midrule
Raw MEA (baseline) & Yes & No & 49.08 & 52.43 & 53.65 & 49.84 \\
Zhao et al. \cite{zhao_deep_2019} & No & Yes (binary spike locations) & N/A\footnotemark[1] & N/A\footnotemark[1] & \begin{tabular}[c]{@{}c@{}}52.13\\($-$2.83)\end{tabular} & N/A\footnotemark[1] \\
FRAME-C V1 & Yes (from spike waveforms) & No & \begin{tabular}[c]{@{}c@{}}58.76\\(+19.72)\end{tabular} & \begin{tabular}[c]{@{}c@{}}59.49\\(+13.47)\end{tabular} & \begin{tabular}[c]{@{}c@{}}57.60\\(+7.36)\end{tabular} & \begin{tabular}[c]{@{}c@{}}59.44\\(+19.26)\end{tabular} \\
FRAME-C V2 & No & Yes & \begin{tabular}[c]{@{}c@{}}59.77\\(+21.78)\end{tabular} & \begin{tabular}[c]{@{}c@{}}61.05\\(+16.44)\end{tabular} & \begin{tabular}[c]{@{}c@{}}55.92\\(+4.23)\end{tabular} & \begin{tabular}[c]{@{}c@{}}61.17\\(+22.74)\end{tabular} \\
FRAME-C V3 & Yes (from spike waveforms) & Yes & \begin{tabular}[c]{@{}c@{}}\textbf{62.31}\\\textbf{(+26.95)}\end{tabular} & \begin{tabular}[c]{@{}c@{}}\textbf{63.49}\\\textbf{(+21.09)}\end{tabular} & \begin{tabular}[c]{@{}c@{}}\textbf{62.58}\\\textbf{(+16.64)}\end{tabular} & \begin{tabular}[c]{@{}c@{}}\textbf{63.08}\\\textbf{(+26.56)}\end{tabular} \\
\botrule
\end{tabular*}
\footnotetext{The effect of using handcrafted features to support automated feature extraction by a deep learning method. To evaluate FRAME-C, experiments were performed with different levels of automated feature extraction and handcrafted features. Accuracies are averaged across all 40 recording days. V1: waveforms only; V2: handcrafted features only; V3: waveforms + handcrafted features. Values are mean test accuracies (\%); values in parentheses are the relative change (\%) with respect to the raw MEA baseline for the same classifier. Bold values indicate the best result for each classifier.}
\footnotetext[1]{Not evaluated, since the pipeline by Zhao et al. was originally proposed with a CNN classifier.}
\end{table}

\begin{table}[htbp]
\caption{Test accuracy of FRAME-C after removing unreliable data samples}\label{tab:results_filtered}
\footnotesize\setlength{\tabcolsep}{3pt}
\begin{tabular*}{\textwidth}{@{\extracolsep\fill}>{\raggedright\arraybackslash}p{2.0cm}>{\raggedright\arraybackslash}p{1.9cm}>{\raggedright\arraybackslash}p{2.2cm}cccc@{}}
\toprule
Classification pipeline & Automated feature extraction & Handcrafted features (domain knowledge) & \multicolumn{4}{@{}c@{}}{Deep learning classifier} \\\cmidrule{4-7}
 & & & Informer & TimesNet & CNN & LSTM \\
\midrule
Raw MEA recording & Yes & No & 53.35 & 50.45 & 53.35 & 50.60 \\
Pipeline by Zhao et al. \cite{zhao_deep_2019} & No & Yes (binary spike locations) & N/A\footnotemark[1] & N/A\footnotemark[1] & \begin{tabular}[c]{@{}c@{}}56.42\\(+5.76)\end{tabular} & N/A\footnotemark[1] \\
FRAME-C V1 & Yes (from spike waveforms) & No & \begin{tabular}[c]{@{}c@{}}57.71\\(+8.17)\end{tabular} & \begin{tabular}[c]{@{}c@{}}62.47\\(+23.83)\end{tabular} & \begin{tabular}[c]{@{}c@{}}63.09\\(+18.26)\end{tabular} & \begin{tabular}[c]{@{}c@{}}62.29\\(+23.10)\end{tabular} \\
FRAME-C V2 & No & Yes & \begin{tabular}[c]{@{}c@{}}64.73\\(+21.33)\end{tabular} & \begin{tabular}[c]{@{}c@{}}65.12\\(+29.08)\end{tabular} & \begin{tabular}[c]{@{}c@{}}\textbf{66.21}\\\textbf{(+24.11)}\end{tabular} & \begin{tabular}[c]{@{}c@{}}66.45\\(+31.32)\end{tabular} \\
FRAME-C V3 & Yes (from spike waveforms) & Yes & \begin{tabular}[c]{@{}c@{}}\textbf{66.87}\\\textbf{(+25.35)}\end{tabular} & \begin{tabular}[c]{@{}c@{}}\textbf{68.80}\\\textbf{(+36.37)}\end{tabular} & \begin{tabular}[c]{@{}c@{}}66.12\\(+23.93)\end{tabular} & \begin{tabular}[c]{@{}c@{}}\textbf{68.14}\\\textbf{(+34.62)}\end{tabular} \\
\botrule
\end{tabular*}
\footnotetext{Unreliable samples (low-activity cultures, Figure \ref{fig:mfr_by_cell_line}) were removed before training and evaluation. V1: waveforms only; V2: handcrafted features only; V3: waveforms + handcrafted features. Values are mean test accuracies (\%); values in parentheses are the relative change (\%) with respect to the raw MEA baseline for the same classifier. Bold values indicate the best result for each classifier.}
\footnotetext[1]{Not evaluated, since the pipeline by Zhao et al. was originally proposed with a CNN classifier.}
\end{table}

\begin{table}[htbp]
\caption{Test accuracy of FRAME-C for the cell lines derived from identical twins}\label{tab:results_twins}
\footnotesize\setlength{\tabcolsep}{3pt}
\begin{tabular*}{\textwidth}{@{\extracolsep\fill}>{\raggedright\arraybackslash}p{2.0cm}>{\raggedright\arraybackslash}p{1.9cm}>{\raggedright\arraybackslash}p{2.2cm}cccc@{}}
\toprule
Classification pipeline & Automated feature extraction & Handcrafted features (domain knowledge) & \multicolumn{4}{@{}c@{}}{Deep learning classifier} \\\cmidrule{4-7}
 & & & Informer & TimesNet & CNN & LSTM \\
\midrule
Raw MEA recording & Yes & No & 57.92 & 57.31 & 53.04 & 50.00 \\
Pipeline by Zhao et al. \cite{zhao_deep_2019} & No & Yes (binary spike locations) & N/A\footnotemark[1] & N/A\footnotemark[1] & \begin{tabular}[c]{@{}c@{}}\textbf{75.62}\\(+42.57)\end{tabular} & N/A\footnotemark[1] \\
FRAME-C V1 & Yes (from spike waveforms) & No & \begin{tabular}[c]{@{}c@{}}\textbf{74.78}\\(+29.11)\end{tabular} & \begin{tabular}[c]{@{}c@{}}72.72\\(+26.89)\end{tabular} & \begin{tabular}[c]{@{}c@{}}71.88\\(+35.52)\end{tabular} & \begin{tabular}[c]{@{}c@{}}74.05\\(+48.10)\end{tabular} \\
FRAME-C V2 & No & Yes & \begin{tabular}[c]{@{}c@{}}64.73\\(+11.76)\end{tabular} & \begin{tabular}[c]{@{}c@{}}71.40\\(+24.59)\end{tabular} & \begin{tabular}[c]{@{}c@{}}69.81\\(+31.62)\end{tabular} & \begin{tabular}[c]{@{}c@{}}71.94\\(+43.88)\end{tabular} \\
FRAME-C V3 & Yes (from spike waveforms) & Yes & \begin{tabular}[c]{@{}c@{}}74.25\\(+28.19)\end{tabular} & \begin{tabular}[c]{@{}c@{}}\textbf{73.85}\\(+28.86)\end{tabular} & \begin{tabular}[c]{@{}c@{}}72.13\\(+35.99)\end{tabular} & \begin{tabular}[c]{@{}c@{}}\textbf{75.08}\\(+50.16)\end{tabular} \\
\botrule
\end{tabular*}
\footnotetext{Models were trained to distinguish the sALS 2 and Control 2 lines, which were derived from identical twins. V1: waveforms only; V2: handcrafted features only; V3: waveforms + handcrafted features. Values are mean test accuracies (\%); values in parentheses are the relative change (\%) with respect to the raw MEA baseline for the same classifier. Bold values indicate the best result for each classifier.}
\footnotetext[1]{Not evaluated, since the pipeline by Zhao et al. was originally proposed with a CNN classifier.}
\end{table}

\textbf{Raw MEA recordings:} In the first experiment, we train each model on raw MEA recordings to establish a baseline classification performance without any handcrafted features for incorporating domain knowledge. Only bandpass filtering and normalization are applied as preprocessing steps. Therefore, we rely on the deep learning model's ability to automatically extract features from the raw MEA signal. As shown in the results, none of the models perform well, with all classifiers achieving just above 50\% accuracy—only marginally better than random classification. The high dimensionality of raw MEA recordings, with relevant information concentrated in spikes, makes it challenging for even state-of-the-art time series classification models to extract useful features without domain-knowledge-informed preprocessing.

\textbf{Pipeline by Zhao et al.:} In the second experiment, we evaluate the pipeline proposed by Zhao et al. \cite{zhao_deep_2019}. This approach uses spike extraction and dimensionality reduction through binning, preserving only the presence and absence of spikes while discarding spike characteristics. Since this pipeline was originally designed for a CNN classifier, we only evaluate it with that model, achieving an accuracy of 56.42\%. Although this method shows an improvement over the baseline, the absence of spike characteristics may limit its ability to capture finer details relevant for classification.

\textbf{FRAME-C V1:} Next, we evaluate the FRAME-C pipeline using only spike waveforms for automated feature extraction. This approach leads to a substantial improvement in classification accuracy across all models, with a peak accuracy of 63.09\%. Compared to raw MEA recordings, the inclusion of spike waveforms enhances model performance by providing temporal information, thus enabling deep learning models to extract more informative features. These results also surpass those obtained with the pipeline proposed by Zhao et al., highlighting the importance of spike characteristics for classification beyond merely detecting spike presence.

\textbf{FRAME-C V2:} We then evaluate the FRAME-C pipeline using only handcrafted features. This configuration incorporates domain knowledge by using handcrafted spike features and extracted spikes that provide temporal information, without performing any automated feature extraction. This approach yields further improvements, achieving gains of 2–3\% over the experiments that used only spike waveforms. These results indicate that while deep learning models can effectively learn from spike waveforms, handcrafted features capture additional relevant information that enhances classification accuracy.

\textbf{FRAME-C V3:} In the final experiment, we evaluate the complete FRAME-C pipeline, which combines both spike waveforms for automated feature extraction and handcrafted spike features. This configuration achieves the highest classification accuracy across all models, demonstrating the complementary nature of the two feature types. Handcrafted features capture key statistical and morphological properties, while deep learning models extract additional patterns from the sequence of spike waveforms. These findings highlight the effectiveness of knowledge augmentation in enhancing class differentiation and underscore the importance of domain-knowledge-driven preprocessing in the FRAME-C pipeline.

\subsection{Performance variation across maturation days}

\begin{figure}[htbp]
  \centering
  \includegraphics[width=\linewidth]{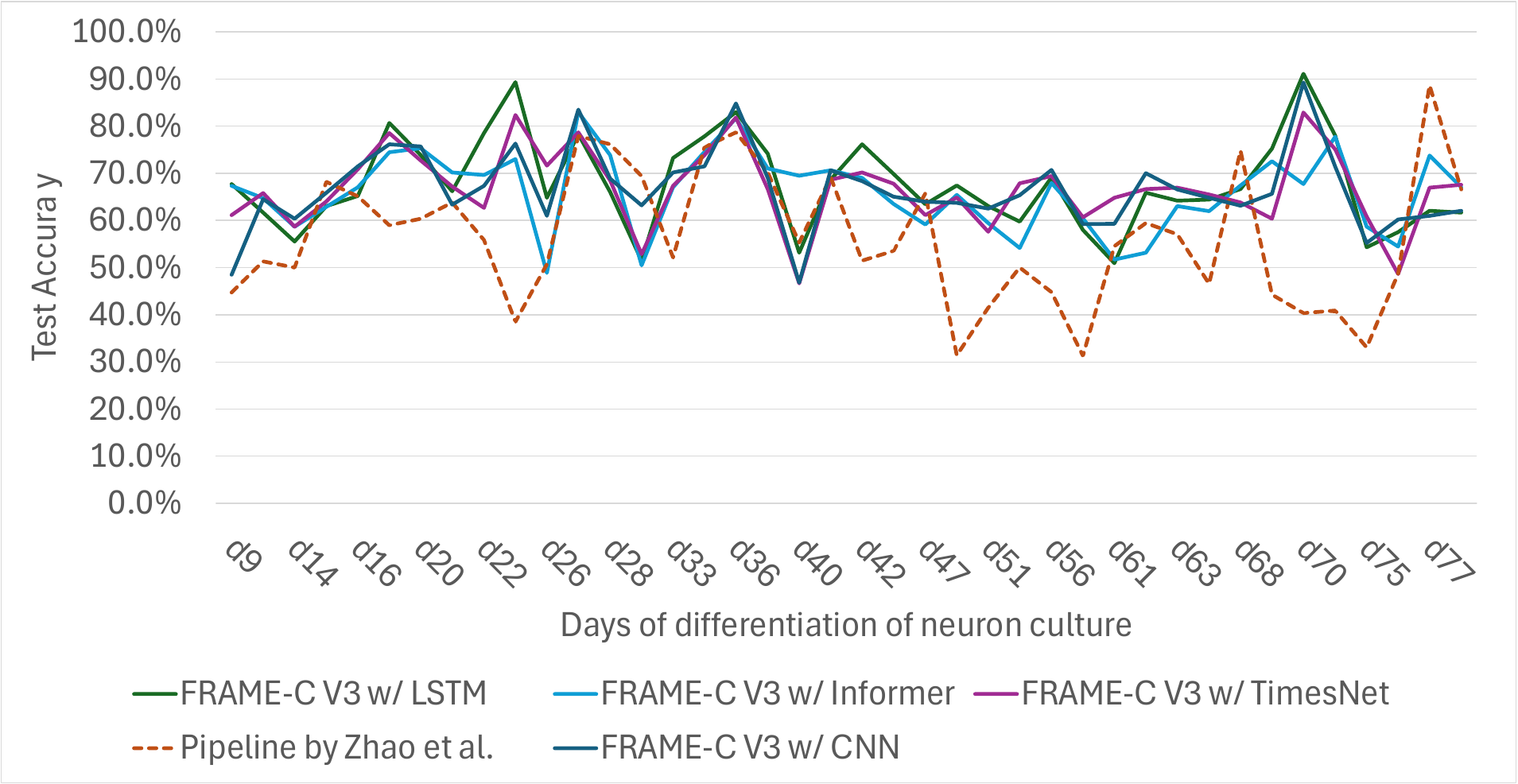}
  \caption{\textbf{Test accuracy of each model across days of neuronal maturation.} The change in test accuracy for each model versus the number of days (d) of neuronal maturation post-differentiation.}
  \label{fig:acc_vs_days}
\end{figure}

Figure \ref{fig:acc_vs_days} illustrates the performance variation of each model across datasets recorded at different maturation days. To maintain clarity and avoid an excessive number of results, only V3, the highest-performing version, is included. FRAME-C based methods demonstrate consistent performance improvements across all maturation stages compared to the pipeline proposed by Zhao et al. 

\subsection{Feature importances}

To further investigate the influence of individual features on model performance, we conducted permutation feature importance experiments across all datasets. These experiments aimed to quantify the contribution of each handcrafted feature by assessing the decrease in model accuracy when a single feature is removed while keeping other features intact. By applying this technique to our ML pipeline, we identified the features that most strongly influenced classification accuracy. We perform the feature importance experiment on both the simulated and real-world MEA datasets. 

For each feature $k$ we train and evaluate FRAME-C with that feature removed. If the test accuracy when feature $k$ is removed is given as $acc_k$, the feature importance $\mathrm{FI}_k$ is given by

\begin{equation}\label{eq:fi}
    \mathrm{FI}_k = acc_{all} - acc_k
\end{equation}

where $acc_{all}$ is the accuracy when trained with all the features. The calculated feature importance scores for the simulated data are given in Table \ref{tab:sim_data_imp} and for the ALS MEA data in Table \ref{tab:real_data_imp}. Since the ALS data and the simulated data do not reflect the same types of cells, we do not expect both of the datasets to yield similar feature importances.

\begin{table}[htbp]
\caption{Importance of each handcrafted feature for classifying simulated MEA recordings}\label{tab:sim_data_imp}
\begin{tabular}{@{}lr@{}}
\toprule
Feature & Permutation feature importance (\%) \\
\midrule
Spike duration & \textbf{23.40} \\
Spike amplitude & 1.60 \\
Inter-spike interval & 0.00 \\
Burst duration & 0.00 \\
Number of spikes per burst & $-$0.50 \\
Burst spike rate & $-$1.00 \\
\botrule
\end{tabular}
\footnotetext{The permutation feature importance is calculated as the drop in test accuracy when the corresponding feature is removed from the dataset (Eq. \ref{eq:fi}). Bold indicates the most important feature.}
\end{table}

\begin{table}[htbp]
\caption{Importance of each handcrafted feature for classifying ALS MEA recordings}\label{tab:real_data_imp}
\begin{tabular}{@{}lr@{}}
\toprule
Feature & Permutation feature importance (\%) \\
\midrule
Inter-spike interval & \textbf{12.30} \\
Spike amplitude & 2.90 \\
Number of spikes per burst & 1.10 \\
Spike duration & 0.80 \\
Burst spike rate & 0.40 \\
Burst duration & $-$1.00 \\
\botrule
\end{tabular}
\footnotetext{The permutation feature importance is calculated as the drop in test accuracy when the corresponding feature is removed from the dataset (Eq. \ref{eq:fi}). Bold indicates the most important feature.}
\end{table}

For the simulated MEA data, spike duration emerged as the most critical feature, with an importance value of 23.40\%, followed by spike amplitude at 1.60\%. In contrast, features such as inter-spike interval and burst duration showed no measurable impact, and some features, like number of spikes per burst and burst spike rate, even showed negative importance, suggesting redundancy or noise. For the ALS MEA data, inter-spike interval was the most influential feature, with an importance of 12.30\%, followed by spike amplitude at 2.90\%. Features like number of spikes per burst and burst spike rate showed minor positive contributions, whereas burst duration had a slight negative importance. 

For both datasets, a notable observation is that the burst related features show little to no importance to the result. We note that all the information in the burst features is inherently available within the spike features. This is because the burst features are directly derived from spike related features. Therefore, we can reasonably assume that the model is successful in implicitly learning any burst related features that are relevant to the classification. Due to this observation, most later experiments were performed with the burst related features removed. 

\begin{figure}[htbp]
    \centering
    \includegraphics[width=\linewidth,height=0.7\textheight,keepaspectratio]{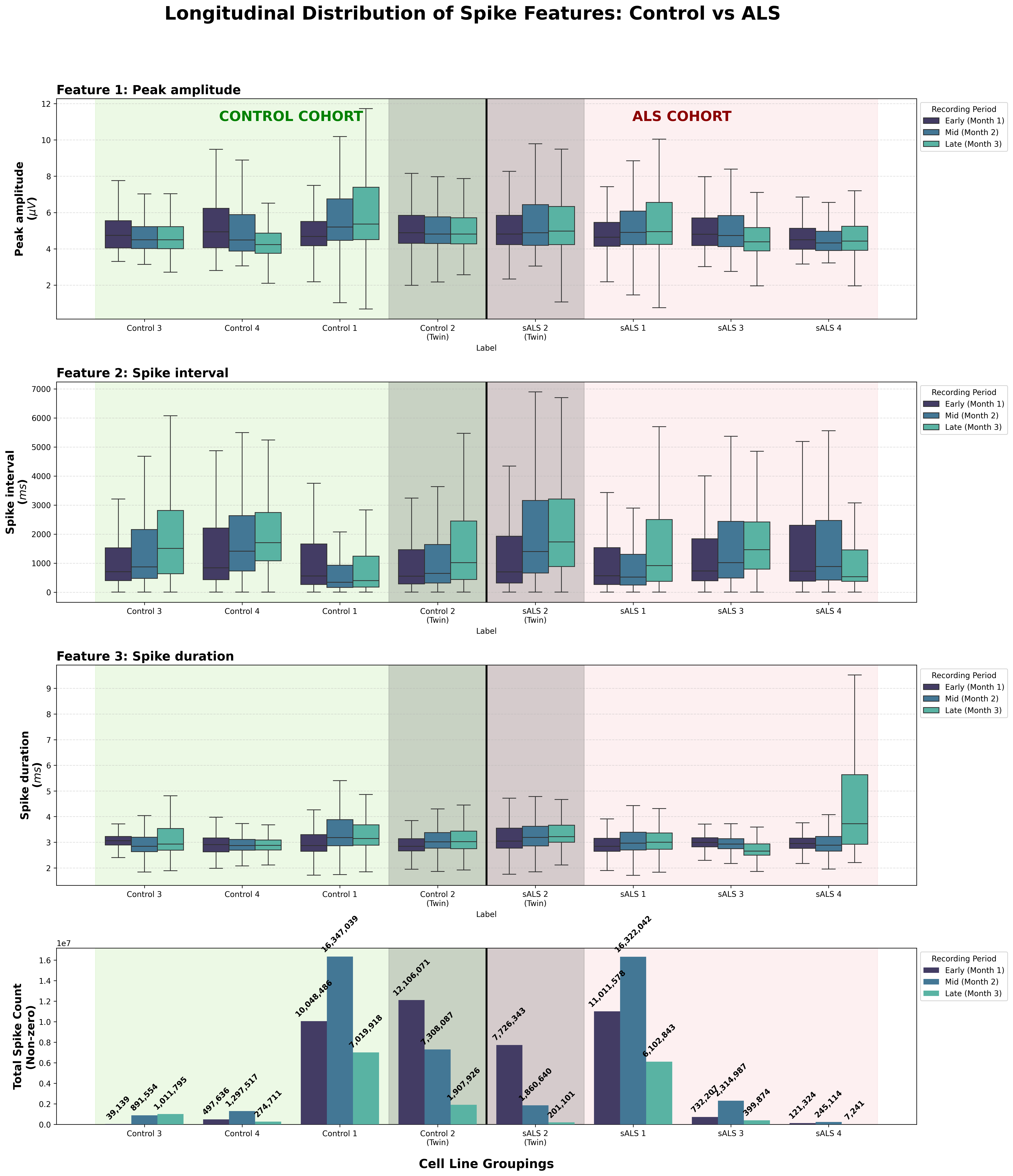}
    \caption{\textbf{Longitudinal distribution of handcrafted features and sample yields across ALS and control lines.} Boxplots illustrate the variance of extracted features for each donor-derived line across the early (month 1), mid (month 2) and late (month 3) recording timepoints. The accompanying bar chart details the total number of non-zero spike samples recorded per line at each respective stage.}
    \label{fig:feature_distributions}
\end{figure}

To evaluate the baseline discriminative power of standard handcrafted features on the real-world human iPSC-derived MEA dataset, we analyzed handcrafted spike features across early, mid, and late recording timepoints (Figure \ref{fig:feature_distributions}). This temporal partitioning was employed not to track disease progression, but to ensure a robust and fair evaluation across the entire recording lifespan. Crucially, to prevent pseudo-replication and accurately reflect true biological and technical variance, individual spike data was aggregated at the well level ($N = 23$--$24$ per cohort) prior to statistical comparison. Mann-Whitney U tests revealed no statistically significant differences between the ALS and control cohorts for peak amplitude, inter-spike interval, or spike duration at any evaluated timepoint ($p > 0.05$, Table \ref{tab:well_aggregated_stats}), with effect sizes remaining consistently small across all metrics (Cliff's $|\delta| < 0.3$). 

\begin{table}[htbp]
\caption{Well-level statistical comparison of handcrafted features between ALS and control lines}\label{tab:well_aggregated_stats}
\footnotesize\setlength{\tabcolsep}{3pt}
\begin{tabular*}{\textwidth}{@{\extracolsep\fill}llccccccc@{}}
\toprule
Feature & Timepoint & $N_{\text{Ctrl}}$ & $N_{\text{ALS}}$ & Median (Ctrl) & Median (ALS) & $p$-value & Sig. & Cliff's $\delta$ \\
\midrule
\multirow{3}{*}{\begin{tabular}[c]{@{}l@{}}Peak amplitude\\($\mu$V)\end{tabular}}
& Early (month 1) & 23 & 23 & 4.714 & 4.596 & 0.1875 & ns & $-$0.229 \\
& Mid (month 2) & 24 & 24 & 4.702 & 4.649 & 0.3808 & ns & $-$0.149 \\
& Late (month 3) & 24 & 24 & 4.669 & 4.573 & 0.5848 & ns & $-$0.094 \\
\midrule
\multirow{3}{*}{\begin{tabular}[c]{@{}l@{}}Spike interval\\(ms)\end{tabular}}
& Early (month 1) & 23 & 23 & 643.0 & 857.5 & 0.2065 & ns & 0.219 \\
& Mid (month 2) & 24 & 24 & 777.0 & 1059.5 & 0.1835 & ns & 0.226 \\
& Late (month 3) & 24 & 24 & 1370.0 & 1142.5 & 0.7650 & ns & $-$0.052 \\
\midrule
\multirow{3}{*}{\begin{tabular}[c]{@{}l@{}}Spike duration\\(ms)\end{tabular}}
& Early (month 1) & 23 & 23 & 2.888 & 2.957 & 0.1663 & ns & 0.240 \\
& Mid (month 2) & 24 & 24 & 3.021 & 2.990 & 0.7029 & ns & $-$0.066 \\
& Late (month 3) & 24 & 24 & 3.025 & 3.173 & 0.0814 & ns & 0.295 \\
\botrule
\end{tabular*}
\footnotetext{Longitudinal statistical analysis of handcrafted features between ALS and control cell lines aggregated at the well level. No statistically significant differences were observed ($p > 0.05$, Mann--Whitney U test) at any timepoint, with small effect sizes (Cliff's $\delta$) across all metrics. ns: not significant.}
\end{table}

While these baseline waveform characteristics confirm overall culture viability, their inability to systematically distinguish between healthy and diseased states underscores the limitations of relying solely on traditional, low-dimensional handcrafted feature analysis of MEA data. However, relying exclusively on automated feature extraction from raw spike waveforms also yields suboptimal classification performance. Neither handcrafted features nor automatically derived latent representations possess the individual discriminative power necessary to reliably predict the ALS disease state in these complex, donor-derived models. This critically motivates the knowledge-augmented architecture of the FRAME-C pipeline. By fusing domain-specific handcrafted features with high-dimensional latent representations extracted via deep learning, FRAME-C is designed to bridge the gap between biological interpretability and automated pattern recognition, providing a unified framework to overcome the limitations of individual feature sets.

\subsection{Effect of knowledge augmentation using spike features}

As noted earlier, a key aspect of FRAME-C is the use of handcrafted features for incorporating domain knowledge into the deep learning pipeline. This approach is motivated by the hypothesis that handcrafted features may explicitly capture useful information, that the deep learning model may not be able to automatically learn. To evaluate the contribution of the spike waveform data and handcrafted features towards the classification performance of FRAME-C, we conducted experiments using three different feature configurations to systematically assess how each feature set affects the performance of each model. For further comparison, we also evaluate the base model's performance when simply trained on the raw MEA recordings, without applying the FRAME-C spike detection and extraction steps. We assessed:

\begin{itemize}
    \item only the handcrafted spike features;
    \item only the raw spike waveforms; and
    \item a combination of both the handcrafted features and raw waveforms.
\end{itemize}

The experimental results, as summarized in Table \ref{tab:results_filtered}, demonstrate the complementary nature of spike features and raw spike waveforms in enhancing model performance. For all four models—Informer, LSTM, TimesNet, and CNN—the combination of spike features and waveforms yielded the highest test accuracy compared to using either feature type individually. Notably, the TimesNet model achieved the best overall accuracy of 68.80\% when trained with both feature types, suggesting its ability to effectively capture temporal dynamics in conjunction with the spatial information provided by the waveform. The spike features alone consistently outperformed the spike waveforms alone across all models, with accuracy improvements ranging from approximately 2\% to 7\%. This indicates that the extracted features contain critical information that is useful for classification. However, the synergy observed when combining the features highlights the added value of incorporating raw waveforms, which likely preserve subtle patterns not captured by the spike features. This balanced contribution is further evidenced by the learnable fusion weights, which consistently converged within the range of 0.45 to 0.55, demonstrating that the model effectively utilized both the raw waveforms and handcrafted features rather than relying on a single data source. These findings emphasize the importance of leveraging both handcrafted features and raw spike waveform data to maximize the accuracy of the machine learning pipeline.

\subsection{Effect of data augmentation}

To further explore the impact of data augmentation by using overlapping time splits, we conducted additional experiments with and without data augmentation. Data augmentation is expected to enhance model performance by providing a more diverse set of training examples. We define the data augmentation factor $\alpha$ as the ratio of the size of the augmented dataset to the size of the original dataset. This can also be expressed in terms of the window size and step size as follows:

\begin{equation}
    \alpha = \frac{\text{augmented data size}}{\text{original data size}} = \frac{\text{window size}}{\text{step size}}
\end{equation}

In our experiments, we use an augmentation factor $\alpha = 10$, meaning the augmented dataset contains 10 times more samples than the original dataset. We train our machine learning pipeline on both the original and augmented datasets, and compare the test accuracies of the two configurations. The results of these experiments are shown in Table \ref{tab:data_aug_results}. We observe a clear positive effect of data augmentation, with all tested methods demonstrating improved performance when trained on the augmented dataset.

\begin{table}[htbp]
\caption{Effect of data augmentation on test accuracy}\label{tab:data_aug_results}
\begin{tabular}{@{}lcc@{}}
\toprule
Method & No data augmentation (\%) & With data augmentation (\%) \\
\midrule
CNN classifier (Zhao et al. \cite{zhao_deep_2019}) & 57.24 & 56.44 \\
FRAME-C V3 with Informer & 63.27 & 65.92 \\
FRAME-C V3 with TimesNet & 65.74 & 66.76 \\
FRAME-C V3 with CNN & 66.28 & 66.66 \\
FRAME-C V3 with LSTM & 65.58 & \textbf{67.82} \\
\botrule
\end{tabular}
\footnotetext{Data augmentation is achieved by splitting each recording into multiple segments with overlapping windows. The overlap is configured to produce 10 times more samples compared to non-overlapping splits. Each method is evaluated based on test accuracy. Bold indicates the highest accuracy.}
\end{table}

\subsection{Performance on simulated data}

To further evaluate the performance of FRAME-C on a different type of dataset, we conducted experiments with FRAME-C to classify simulated MEA recordings. We use data simulated from L5 cortical neuronal models using MEArec \cite{buccino_mearec_2021}, as outlined in Section \ref{sec:dataset}. The simulated recordings are generated using both inhibitory and excitatory cell models, and the models were trained to classify between these two classes of cells. To better mimic real-world data, Gaussian additive noise with a standard deviation of 20\,$\mu$V was added to the simulated recordings.

Multiple datasets of varying sizes were generated, and experiments were repeated on each dataset. For each experiment, we trained both our ML pipeline and the CNN-based method proposed by Zhao et al. \cite{zhao_deep_2019} and compared their test set accuracies. Since FRAME-C with LSTM performed the best in classifying the real-world ALS data, we use the LSTM classifier in these experiments. Moreover, we run the experiments using the highest performing version of the proposed pipeline, FRAME-C V3. To evaluate how each method performs on datasets of different sizes, multiple experiments were performed with simulated datasets consisting of 2, 4, 8, and 16 simulated MEA recordings (each recording is 300 s long). Each recording would correspond to a single well in a multi-well MEA experiment. The results are shown in Table \ref{tab:sim_data_results}.

\begin{table}[htbp]
\caption{Test accuracy on simulated MEA recordings for different dataset sizes}\label{tab:sim_data_results}
\begin{tabular}{@{}ccc@{}}
\toprule
\begin{tabular}[c]{@{}c@{}}Simulated wells (recordings)\\per cell type\end{tabular} & Pipeline by Zhao et al. (\%) & FRAME-C V3 with LSTM (\%) \\
\midrule
2 & 51.12 & 75.06 (+46.83) \\
4 & 75.44 & 82.05 (+8.76) \\
8 & 79.13 & 84.58 (+6.89) \\
16 & 94.07 & \textbf{94.21} (\textbf{+0.15}) \\
\botrule
\end{tabular}
\footnotetext{Performance comparison of FRAME-C V3 with LSTM and the pipeline by Zhao et al. \cite{zhao_deep_2019} on simulated MEA recordings. Dataset size is measured by the number of simulated wells (recordings) per cell type. Values in parentheses are the relative improvement (\%) of FRAME-C over the pipeline by Zhao et al.}
\end{table}

The results demonstrate that the proposed ML pipeline effectively distinguishes between cell types, consistently outperforming the CNN-based method by Zhao et al. This is more evident with smaller datasets, with a relative improvement of over 46\% when trained on just two recordings, highlighting FRAME-C’s ability to learn important phenotypes from the MEA recordings with limited and noisy data.

\section{Conclusion}
\label{sec:conclusion}

In this work, we introduce FRAME-C, a machine learning pipeline that combines domain knowledge with raw signal data to classify MEA recordings across different applications. FRAME-C uses a spike detection algorithm to extract spike waveforms from the MEA recording, which are then fed into the deep learning model for automated feature extraction. Handcrafted features are used to incorporate domain knowledge into the pipeline, further improving classification performance.

We evaluated FRAME-C on MEA data recorded from human iPSC-derived neuronal cultures and its ability to identify electrophysiological phenotypes in ALS by training it to classify ALS-diseased and healthy cell lines. FRAME-C is shown to outperform existing methods, achieving up to a 36.37\% relative improvement in test accuracy.

To further assess the robustness of FRAME-C, we evaluated its performance on simulated MEA data generated from cortical neuronal models. These experiments demonstrate FRAME-C's ability to generalize to a different type of dataset while maintaining strong classification performance. In this application, FRAME-C was trained to classify between inhibitory and excitatory neuronal subtypes—a fundamentally different measurement from the ALS vs control classification—demonstrating that FRAME-C provides accurate classification across diverse electrophysiological phenotype detection tasks. Notably, FRAME-C outperformed the CNN-based approach by Zhao et al., particularly on smaller datasets, where it achieved a relative improvement of over 46\% when trained on just two simulated recordings. With larger simulated datasets, FRAME-C achieved an accuracy of up to 94\%. This highlights FRAME-C's capability to extract meaningful phenotypic differences even from limited and noisy data.

Feature importance experiments reveal that the most discriminative feature varies depending on the dataset: spike duration is the most critical phenotype in classifying the simulated MEA data (distinguishing inhibitory vs excitatory cells), whereas inter-spike interval (ISI) plays the most important role in ALS classification. By leveraging domain-specific knowledge augmentation and deep learning architectures, FRAME-C not only enhances classification accuracy but also provides interpretability by highlighting key phenotypic features.

Although FRAME-C demonstrates strong performance with both \textit{in vitro} and simulated multi-electrode array data, its generalizability to other types of electrophysiological data such as EEG remains untested. Additionally, while the well-wise approach mitigates the misrepresentation of model performance caused by batch effects, the potential influence of such effects on overall performance cannot be ruled out. Future work may explore batch correction methods to further enhance the robustness and accuracy of FRAME-C.

\backmatter


\bmhead{Abbreviations}\mbox{}\par\noindent\begin{tabular}{@{}lp{0.75\textwidth}@{}}
ALS & Amyotrophic lateral sclerosis\\
AP & Action potential\\
BDNF & Brain-derived neurotrophic factor\\
BSR & Burst spike rate\\
CNN & Convolutional neural network\\
DB-cAMP & Dibutyryl cyclic adenosine monophosphate\\
EAP & Extracellular action potential\\
ECG & Electrocardiography\\
EEG & Electroencephalography\\
EMG & Electromyography\\
fALS & Familial amyotrophic lateral sclerosis\\
FRAME-C & Feature-based Raw data Augmentation for Multi-electrode array recorded Electrophysiological signal Classification\\
GDNF & Glial cell line-derived neurotrophic factor\\
IGF & Insulin-like growth factor\\
iPSC & Induced pluripotent stem cell\\
ISI & Inter-spike interval\\
L-AA & L-ascorbic acid\\
LSTM & Long short-term memory\\
MEA & Multi-electrode array\\
MFR & Mean firing rate\\
ML & Machine learning\\
MNP & Motor neuron progenitor\\
NEAA & Non-essential amino acids\\
PBS & Phosphate-buffered saline\\
RNN & Recurrent neural network\\
sALS & Sporadic amyotrophic lateral sclerosis\\
UMAP & Uniform Manifold Approximation and Projection\\
\end{tabular}

\section*{Declarations}

\bmhead{Ethics approval and consent to participate}
The study was conducted according to the National Statement on Ethical Conduct in Human Research and approved by the University of Wollongong Human Research Ethics Committee (approval numbers 13/272 and 19/380). 



\bmhead{Availability of data and materials}
The datasets used and/or analysed during the current study are available from the corresponding author on reasonable request. The code is provided as supplementary.

\bmhead{Competing interests}
The authors declare that they have no competing interests.

\bmhead{Funding}
This research was funded by a FightMND Impact grant (07) awarded to LO and SH and a US Department of Defense Therapeutic Idea Award (AL210047; W81XWH-22-1-0311) awarded to LO and SH. Parts of this research were supported by the Australian Research Council Centre of Excellence in Quantum Biotechnology (CE230100021) awarded to LO and Australian Research Council 
Discovery Project funding (DP220101035) awarded to SH. 


\bmhead{Authors' contributions}
Conceptualisation: NR, DD-H, CG, SM, TM, SS, LO, SH; Methodology: NR, CG, DD-H, SM, TM, SS, LO, SH; Data curation: NR, DD-H; Writing – review \& editing: NR, CG, DD-H, SM, TM, SS, LO, SH; Supervision: LO, SH; Funding acquisition: LO, SH. All authors read and approved the final manuscript.


\bibliography{references}

\begin{thebibliography}{44}
\providecommand{\natexlab}[1]{#1}
\providecommand{\url}[1]{{#1}}
\providecommand{\urlprefix}{URL }
\providecommand{\doi}[1]{\url{https://doi.org/#1}}
\providecommand{\eprint}[2][]{\url{#2}}
 \bibcommenthead

\bibitem[{Al-Zaiti et~al.(2023)Al-Zaiti, Martin-Gill, Zègre-Hemsey, Bouzid, Faramand, Alrawashdeh, Gregg, Helman, Riek, Kraevsky-Phillips, Clermont, Akcakaya, Sereika, Van~Dam, Smith, Birnbaum, Saba, Sejdic, and Callaway}]{al-zaiti_machine_2023}
Al-Zaiti SS, Martin-Gill C, Zègre-Hemsey JK, et~al (2023) Machine learning for {ECG} diagnosis and risk stratification of occlusion myocardial infarction. Nature Medicine 29(7):1804--1813. \doi{10.1038/s41591-023-02396-3}

\bibitem[{Barak(2017)}]{barak_recurrent_2017}
Barak O (2017) Recurrent neural networks as versatile tools of neuroscience research. Current Opinion in Neurobiology 46:1--6. \doi{10.1016/j.conb.2017.06.003}

\bibitem[{Bogahawatte et~al.(2026)Bogahawatte, Seneviratne, Maksour, Do-Ha, Qian, Brennan, Keon, Ooi, and Halgamuge}]{bogahawatte2026model}
Bogahawatte J, Seneviratne S, Maksour S, et~al (2026) Model-agnostic knowledge augmented machine learning for weakly annotated bioimaging data. arXiv preprint arXiv:2610{\char46}13637

\bibitem[{Brown and Al-Chalabi(2017)}]{brown_amyotrophic_2017}
Brown RH, Al-Chalabi A (2017) Amyotrophic lateral sclerosis. New England Journal of Medicine 377(2):162--172

\bibitem[{Bryson et~al.(2022)Bryson, Mendis, Morrisroe, Reid, Halgamuge, and Petrou}]{bryson_classification_2022}
Bryson A, Mendis D, Morrisroe E, et~al (2022) Classification of antiseizure drugs in cultured neuronal networks using multielectrode arrays and unsupervised learning. Epilepsia 63(7):1693--1703. \doi{10.1111/epi.17268}

\bibitem[{Buccino and Einevoll(2021)}]{buccino_mearec_2021}
Buccino AP, Einevoll GT (2021) {MEArec}: {A} {Fast} and {Customizable} {Testbench} {Simulator} for {Ground}-truth {Extracellular} {Spiking} {Activity}. Neuroinformatics 19(1):185--204. \doi{10.1007/s12021-020-09467-7}

\bibitem[{Buccino et~al.(2018)Buccino, Kordovan, Ness, Merkt, Häfliger, Fyhn, Cauwenberghs, Rotter, and Einevoll}]{buccino_combining_2018}
Buccino AP, Kordovan M, Ness TV, et~al (2018) Combining biophysical modeling and deep learning for multielectrode array neuron localization and classification. Journal of Neurophysiology 120(3):1212--1232. \doi{10.1152/jn.00210.2018}

\bibitem[{Buest{\'a}n-Andrade et~al.(2023)Buest{\'a}n-Andrade, Santos, Sierra-Garc{\'\i}a, and Pazmi{\~n}o-Piedra}]{buestan-andrade_comparison_2023}
Buest{\'a}n-Andrade PA, Santos M, Sierra-Garc{\'\i}a JE, et~al (2023) Comparison of lstm, gru and transformer neural network architecture for prediction of wind turbine variables. In: International conference on soft computing models in industrial and environmental applications, Springer, pp 334--343

\bibitem[{Buskila et~al.(2019)Buskila, K{\'e}kesi, Bellot-Saez, Seah, Berg, Trpceski, Yerbury, and Ooi}]{buskila_dynamic_2019}
Buskila Y, K{\'e}kesi O, Bellot-Saez A, et~al (2019) Dynamic interplay between h-current and m-current controls motoneuron hyperexcitability in amyotrophic lateral sclerosis. Cell death \& disease 10(4):310

\bibitem[{Carnevale and Hines(2006)}]{carnevale_neuron_2006}
Carnevale NT, Hines ML (2006) The {NEURON} {Book}. Cambridge University Press, \doi{10.1017/CBO9780511541612}

\bibitem[{Chen et~al.(2013)Chen, Sayana, Zhang, and Le}]{chen_genetics_2013}
Chen S, Sayana P, Zhang X, et~al (2013) Genetics of amyotrophic lateral sclerosis: an update. Molecular Neurodegeneration 8(1):28. \doi{10.1186/1750-1326-8-28}

\bibitem[{Cui et~al.(2023)Cui, Gao, Talamadupula, and Ji}]{cui_knowledge-augmented_2023}
Cui Z, Gao T, Talamadupula K, et~al (2023) Knowledge-{Augmented} {Deep} {Learning} and {Its} {Applications}: {A} {Survey}. IEEE Transactions on Neural Networks and Learning Systems pp 1--21. \doi{10.1109/TNNLS.2023.3338619}

\bibitem[{DeJesus-Hernandez et~al.(2011)DeJesus-Hernandez, Mackenzie, Boeve, Boxer, Baker, Rutherford, Nicholson, Finch, Flynn, Adamson, Kouri, Wojtas, Sengdy, Hsiung, Karydas, Seeley, Josephs, Coppola, Geschwind, Wszolek, Feldman, Knopman, Petersen, Miller, Dickson, Boylan, Graff-Radford, and Rademakers}]{dejesus-hernandez_expanded_2011}
DeJesus-Hernandez M, Mackenzie IR, Boeve BF, et~al (2011) Expanded {GGGGCC} {Hexanucleotide} {Repeat} in {Noncoding} {Region} of {C9ORF72} {Causes} {Chromosome} 9p-{Linked} {FTD} and {ALS}. Neuron 72(2):245--256. \doi{10.1016/j.neuron.2011.09.011}

\bibitem[{Do-Ha et~al.(2018)Do-Ha, Buskila, and Ooi}]{do-ha_impairments_2018}
Do-Ha D, Buskila Y, Ooi L (2018) Impairments in {Motor} {Neurons}, {Interneurons} and {Astrocytes} {Contribute} to {Hyperexcitability} in {ALS}: {Underlying} {Mechanisms} and {Paths} to {Therapy}. Molecular Neurobiology 55(2):1410--1418. \doi{10.1007/s12035-017-0392-y}

\bibitem[{Dong et~al.(2021)Dong, Wang, and Abbas}]{dong_survey_2021}
Dong S, Wang P, Abbas K (2021) A survey on deep learning and its applications. Computer Science Review 40:100379. \doi{10.1016/j.cosrev.2021.100379}

\bibitem[{Dosovitskiy(2020)}]{dosovitskiy_image_2020}
Dosovitskiy A (2020) An image is worth 16x16 words: Transformers for image recognition at scale. arXiv preprint arXiv:2010{\char46}11929

\bibitem[{Gamage et~al.(2026)Gamage, Xia, Rupasinghe, Senevirathne, Senanayake, Malepathirana, Hevapathige, Corbett, O’Brien, Petrou et~al.}]{gamage_knowledge_2026}
Gamage CJ, Xia Y, Rupasinghe R, et~al (2026) Knowledge inclusive machine learning for disease gene prioritisation. bioRxiv pp 2026--04

\bibitem[{Hosseini et~al.(2021)Hosseini, Hosseini, and Ahi}]{hosseini_review_2021}
Hosseini MP, Hosseini A, Ahi K (2021) A {Review} on {Machine} {Learning} for {EEG} {Signal} {Processing} in {Bioengineering}. IEEE reviews in biomedical engineering 14:204--218. \doi{10.1109/RBME.2020.2969915}

\bibitem[{Keles et~al.(2022)Keles, Wijewardena, and Hegde}]{keles_computational_2022}
Keles FD, Wijewardena PM, Hegde C (2022) On {The} {Computational} {Complexity} of {Self}-{Attention}. \doi{10.48550/arXiv.2209.04881}

\bibitem[{Li et~al.(2024)Li, Yang, Lou, and Zhang}]{li_temporal-spectral_2024}
Li R, Yang X, Lou J, et~al (2024) A temporal-spectral graph convolutional neural network model for eeg emotion recognition within and across subjects. Brain Informatics 11(1):30

\bibitem[{Lind{\'e}n et~al.(2014)Lind{\'e}n, Hagen, {\L}{\k{e}}ski, Norheim, Pettersen, and Einevoll}]{linden_lfpy_2014}
Lind{\'e}n H, Hagen E, {\L}{\k{e}}ski S, et~al (2014) Lfpy: a tool for biophysical simulation of extracellular potentials generated by detailed model neurons. Frontiers in neuroinformatics 7:41

\bibitem[{Maksour et~al.(2024)Maksour, Finol-Urdaneta, Hulme, Cabral-da Silva, Targa Dias~Anastacio, Balez, Berg, Turner, Sanz~Mu{\~n}oz, Engel et~al.}]{maksour_alzheimers_2024}
Maksour S, Finol-Urdaneta RK, Hulme AJ, et~al (2024) Alzheimer’s disease induced neurons bearing psen1 mutations exhibit reduced excitability. Frontiers in Cellular Neuroscience 18:1406970

\bibitem[{Malepathirana et~al.(2024)Malepathirana, Senanayake, Gautam, Engel, Balez, Lovelace, Sundaram, Heng, Chow, Marquis, Guillemin, Brew, Jagadish, Ooi, and Halgamuge}]{malepathirana_visualization_2024}
Malepathirana T, Senanayake D, Gautam V, et~al (2024) Visualization of incrementally learned projection trajectories for longitudinal data. Scientific Reports 14(1):13558. \doi{10.1038/s41598-024-63511-z}

\bibitem[{McInnes et~al.(2018)McInnes, Healy, and Melville}]{mcinnes_umap_2018}
McInnes L, Healy J, Melville J (2018) Umap: Uniform manifold approximation and projection for dimension reduction. arXiv preprint arXiv:1802{\char46}03426

\bibitem[{Mendis et~al.(2016)Mendis, Morrisroe, Reid, Halgamuge, and Petrou}]{mendis_use_2016}
Mendis GDC, Morrisroe E, Reid CA, et~al (2016) Use of local field potentials of dissociated cultures grown on multi-electrode arrays for pharmacological assays. In: 2016 38th Annual International Conference of the IEEE Engineering in Medicine and Biology Society (EMBC), IEEE, pp 952--956

\bibitem[{Mossink et~al.(2021)Mossink, Verboven, van Hugte, Klein~Gunnewiek, Parodi, Linda, Schoenmaker, Kleefstra, Kozicz, van Bokhoven, Schubert, Nadif~Kasri, and Frega}]{mossink_human_2021}
Mossink B, Verboven AH, van Hugte EJ, et~al (2021) Human neuronal networks on micro-electrode arrays are a highly robust tool to study disease-specific genotype-phenotype correlations in vitro. Stem Cell Reports 16(9):2182--2196. \doi{10.1016/j.stemcr.2021.07.001}

\bibitem[{Ramaswamy et~al.(2015)Ramaswamy, Courcol, Abdellah, Adaszewski, Antille, Arsever, Atenekeng, Bilgili, Brukau, Chalimourda et~al.}]{ramaswamy_neocortical_2015}
Ramaswamy S, Courcol JD, Abdellah M, et~al (2015) The neocortical microcircuit collaboration portal: a resource for rat somatosensory cortex. Frontiers in neural circuits 9:44

\bibitem[{Renton et~al.(2011)Renton, Majounie, Waite, Simón-Sánchez, Rollinson, Gibbs, Schymick, Laaksovirta, van Swieten, Myllykangas, Kalimo, Paetau, Abramzon, Remes, Kaganovich, Scholz, Duckworth, Ding, Harmer, Hernandez, Johnson, Mok, Ryten, Trabzuni, Guerreiro, Orrell, Neal, Murray, Pearson, Jansen, Sondervan, Seelaar, Blake, Young, Halliwell, Callister, Toulson, Richardson, Gerhard, Snowden, Mann, Neary, Nalls, Peuralinna, Jansson, Isoviita, Kaivorinne, Hölttä-Vuori, Ikonen, Sulkava, Benatar, Wuu, Chiò, Restagno, Borghero, Sabatelli, Heckerman, Rogaeva, Zinman, Rothstein, Sendtner, Drepper, Eichler, Alkan, Abdullaev, Pack, Dutra, Pak, Hardy, Singleton, Williams, Heutink, Pickering-Brown, Morris, Tienari, and Traynor}]{renton_hexanucleotide_2011}
Renton AE, Majounie E, Waite A, et~al (2011) A hexanucleotide repeat expansion in {C9ORF72} is the cause of chromosome 9p21-linked {ALS}-{FTD}. Neuron 72(2):257--268. \doi{10.1016/j.neuron.2011.09.010}

\bibitem[{Robinson et~al.(2026)Robinson, Do-Ha, Cheng, Stevens, Rosa~Porto, Coles, Subachandran, Kalajdzic, Lui, Balez, Sanz~Mu{\~n}oz, Cabral-da Silva, Berg, Morsch, Lisowski, Tan, Burgio, Karl, Lee, Chung, Blair, and Ooi}]{robinson_als-ftd-linked_2026}
Robinson L, Do-Ha D, Cheng F, et~al (2026) {ALS}-{FTD}-linked {CCNF}$^{S621G}$ drives increased hippocampal astrocyte ramification and mitochondrial dysfunction and impairs motor neuron excitability. Journal of Neuroinflammation 23(1):208. \doi{10.1186/s12974-026-03827-x}

\bibitem[{Rosen et~al.(1993)Rosen, Siddique, Patterson, Figlewicz, Sapp, Hentati, Donaldson, Goto, O'Regan, Deng et~al.}]{rosen_mutations_1993}
Rosen DR, Siddique T, Patterson D, et~al (1993) Mutations in cu/zn superoxide dismutase gene are associated with familial amyotrophic lateral sclerosis. Nature 362(6415):59--62

\bibitem[{Sreedharan et~al.(2008)Sreedharan, Blair, Tripathi, Hu, Vance, Rogelj, Ackerley, Durnall, Williams, Buratti et~al.}]{sreedharan_tdp-43_2008}
Sreedharan J, Blair IP, Tripathi VB, et~al (2008) Tdp-43 mutations in familial and sporadic amyotrophic lateral sclerosis. Science 319(5870):1668--1672

\bibitem[{Takahashi and Yamanaka(2006)}]{takahashi_induction_2006}
Takahashi K, Yamanaka S (2006) Induction of {Pluripotent} {Stem} {Cells} from {Mouse} {Embryonic} and {Adult} {Fibroblast} {Cultures} by {Defined} {Factors}. Cell 126(4):663--676. \doi{10.1016/j.cell.2006.07.024}

\bibitem[{Takahashi and Yamanaka(2016)}]{takahashi_decade_2016}
Takahashi K, Yamanaka S (2016) A decade of transcription factor-mediated reprogramming to pluripotency. Nature Reviews Molecular Cell Biology 17(3):183--193. \doi{10.1038/nrm.2016.8}

\bibitem[{Vaswani et~al.(2017)Vaswani, Shazeer, Parmar, Uszkoreit, Jones, Gomez, Kaiser, and Polosukhin}]{vaswani_attention_2017}
Vaswani A, Shazeer N, Parmar N, et~al (2017) Attention is all you need. Advances in neural information processing systems 30

\bibitem[{Viswan et~al.(2026)Viswan, Shaffi, Malathy, Chemmalar~Selvi, Kavitha, Abdesselam, Wang, Suganthan, Shezawi, and Mahmud}]{viswan_multimodal_2026}
Viswan V, Shaffi N, Malathy E, et~al (2026) Multimodal fusion and explainability of artificial intelligence models in alzheimer’s disease detection. Brain Informatics 13(1):5

\bibitem[{Wainger et~al.(2014)Wainger, Kiskinis, Mellin, Wiskow, Han, Sandoe, Perez, Williams, Lee, Boulting, Berry, Brown, Cudkowicz, Bean, Eggan, and Woolf}]{wainger_intrinsic_2014}
Wainger BJ, Kiskinis E, Mellin C, et~al (2014) Intrinsic membrane hyperexcitability of amyotrophic lateral sclerosis patient-derived motor neurons. Cell Reports 7(1):1--11. \doi{10.1016/j.celrep.2014.03.019}

\bibitem[{Wang et~al.(2009)Wang, Zhang, Zhou, Barakat, and Querfurth}]{wang_opposite_2009}
Wang Y, Zhang G, Zhou H, et~al (2009) Opposite effects of low and high doses of a$\beta$42 on electrical network and neuronal excitability in the rat prefrontal cortex. PLoS One 4(12):e8366

\bibitem[{Wang et~al.(2024)Wang, Wu, Dong, Liu, Long, and Wang}]{wang_deep_2024}
Wang Y, Wu H, Dong J, et~al (2024) Deep {Time} {Series} {Models}: {A} {Comprehensive} {Survey} and {Benchmark}. \doi{10.48550/arXiv.2407.13278}

\bibitem[{Weir et~al.(2015)Weir, Blanquie, Kilb, Luhmann, and Sinning}]{weir_comparison_2015}
Weir K, Blanquie O, Kilb W, et~al (2015) Comparison of spike parameters from optically identified {GABAergic} and glutamatergic neurons in sparse cortical cultures. Frontiers in Cellular Neuroscience 8:460. \doi{10.3389/fncel.2014.00460}

\bibitem[{Wu et~al.(2022)Wu, Hu, Liu, Zhou, Wang, and Long}]{wu_timesnet_2022}
Wu H, Hu T, Liu Y, et~al (2022) Timesnet: Temporal 2d-variation modeling for general time series analysis. arXiv preprint arXiv:2210{\char46}02186

\bibitem[{Xu et~al.(2025)Xu, Chen, Deng, Shen, Wei, Yu, and Zhang}]{xu_enhancing_2025}
Xu X, Chen N, Deng Y, et~al (2025) Enhancing neural signal quality: a spike-denoising model with bilstm and attention mechanism. Brain Informatics 12(1):28

\bibitem[{Yousif et~al.(2019)Yousif, Zakaria, Rahim, Salleh, Mahmood, Alfarhan, Kamarudin, Mamduh, Hasan, and Hussain}]{yousif_assessment_2019}
Yousif HA, Zakaria A, Rahim NA, et~al (2019) Assessment of {Muscles} {Fatigue} {Based} on {Surface} {EMG} {Signals} {Using} {Machine} {Learning} and {Statistical} {Approaches}: {A} {Review}. IOP Conference Series: Materials Science and Engineering 705(1):012010. \doi{10.1088/1757-899X/705/1/012010}

\bibitem[{Zhao et~al.(2019)Zhao, Guzman, Audouard, Cheng, Hansma, Kosik, and Petzold}]{zhao_deep_2019}
Zhao Y, Guzman E, Audouard M, et~al (2019) A deep learning framework for classification of in vitro multi-electrode array recordings. arXiv preprint arXiv:1906{\char46}02241

\bibitem[{Zhou et~al.(2021)Zhou, Zhang, Peng, Zhang, Li, Xiong, and Zhang}]{zhou_informer_2021}
Zhou H, Zhang S, Peng J, et~al (2021) Informer: {Beyond} {Efficient} {Transformer} for {Long} {Sequence} {Time}-{Series} {Forecasting}. Proceedings of the AAAI Conference on Artificial Intelligence 35(12):11106--11115. \doi{10.1609/aaai.v35i12.17325}

\end{thebibliography}

\end{document}